\documentclass[a4paper]{article}

\usepackage[english]{babel}
\usepackage[utf8x]{inputenc}
\usepackage[T1]{fontenc}

\usepackage[a4paper,top=3cm,bottom=2cm,left=3cm,right=3cm,marginparwidth=1.75cm]{geometry}
\usepackage{authblk}

\usepackage{amsmath}
\usepackage{graphicx}
\usepackage[colorinlistoftodos]{todonotes}
\usepackage[colorlinks=true, allcolors=blue]{hyperref}
\usepackage[natbibapa]{apacite}
\usepackage{psboxit}

\usepackage{cases}

\title{On how traffic signals impact the fundamental diagrams of urban roads}

\author[1]{Chao Zhang\thanks{Corresponding author: chaozhng@google.com}}
\author[1]{Yechen Li}
\author[1]{Neha Arora}
\author[1,2]{Carolina Osorio}

\affil[1]{Google Research, Mountain View, CA, USA}
\affil[2]{HEC Montr\'{e}al, Montr\'{e}al, QC, Canada}
\date{\vspace{-5ex}}
\begin{document}
\maketitle
\renewcommand{\abstractname}{SHORT SUMMARY}

\begin{abstract}
%% Text of abstract
Being widely adopted by the transportation and planning practitioners, the fundamental diagram (FD) is the primary tool used to relate the key macroscopic traffic variables of speed, flow, and density. We empirically analyze the relation between vehicular space-mean speeds and flows given different signal settings and postulate a parsimonious parametric function form of the traditional FD where its function parameters are explicitly modeled as a function of the signal plan factors. We validate the proposed formulation using data from signalized urban road segments in Salt Lake City, Utah, USA. The proposed formulation builds our understanding of how changes to signal settings impact the FDs, and more generally the congestion patterns, of signalized urban segments. 

% This can help inform enhanced traffic management strategies. More generally, it enables many transportation planning and management applications whose operations heavily depend on the availability of FDs.
\end{abstract}


\section{Introduction}
Fundamental diagrams (FDs) are recognized to be an important tool in the design and operations of segments (i.e., roadways) with uninterrupted traffic, such as  highways. Since Greenshields' seminal work on freeway FDs \citep{greenshields35, kuhne08}, there have been various efforts in the research community to expand it to urban roads with interrupted traffic \citep{brockfeld08} and to elucidate certain functional forms for both highways and urban settings \citep{hall92, wu11, dakic18}. Recent work by \cite{li22} addresses this challenge by deriving a parametric form for FDs that are associated with urban signalized road segments. As an indispensable part of the urban transportation infrastructure, traffic lights are arguably one of the most critical factors that are related to urban congestion. The work of \cite{li22}  showed that knowledge of the signal setting can help explain the shape of the FD of segments that are upstream of the signalized segment. However, the parameters of the FD of the signalized segment are estimated numerically through field data. They lack interpretability, and hence do not provide a physical explanation into what aspects of traffic supply govern the shape of the FD. This paper addresses this challenge. It considers signalized urban segments. It extends the work of \cite{li22} by formulating the parameters as explicit functions of the signal settings. 
% Specifically, this paper investigates an explicit relationship between the underlying FDs and the corresponding traffic infrastructure, e.g., signal controls, which are broadly directly available or observable. This paper presents a mathematical expression for urban FDs whose hyper-parameters are highly interpretable and can be fully determined by the associated traffic signal plan factors, e.g., green split. 
As a result, this work relates the build-up and dissipation of congestion to prevailing signal settings. 

% The proposed formulation can help enhance various traffic control and transportation planning techniques, where FDs are readily used. 
%For instance, this will be relevant to the enhancement of routing algorithms in Google Maps \citep{google_map_guidance}. 

The major contributions of this paper are as follows: (1) the formulation provides insights into the interplay of traffic signal control and urban congestion formation and propagation; and (2) the formulation facilitates the scalability of the urban FD by reducing the reliance on segment-specific speed and flow data.

\section{Methodology}
We first briefly review the urban FD work of \cite{li22}, that the proposed methodology in this paper extends. Then, we present the proposed formulation.

\subsection{Urban FD for signalized segments}
Inspired by the commonly used FDs for segments with uninterrupted traffic of \cite{may67}, a parsimonious parametric function is postulated for the FDs of urban signalized segments. The formulation relates vehicular space-mean speeds and flows:
\begin{equation}
    v = v_{max} \left( 1- \left(\frac{q}{q_{cap}} \right)^\alpha \right)^\beta, 
    \label{eq:FD1}  
\end{equation}

\noindent where the space-mean speed is denoted by $v$, the speed limit is denoted by $v_{max}$, the mean flow per lane is denoted by $q$, and the flow capacity (per lane) is denoted by $q_{cap}$. The parameters $\alpha$ and $\beta$ are the parameters that are numerically estimated and lack a physical interpretation. This paper proposes a formulation of these two parameters that is interpretable and is directly related to the signal settings. 

Equation~\eqref{eq:FD1} takes a similar functional form as that of \cite{may67}. However, it relates speed and flow, while \cite{may67} relates speed and density. A discussion on the limitations and open questions related to the FD of Equation~\eqref{eq:FD1} is given in \cite{li22}.

\subsection{Signal-parametrized FD}
In this paper, we consider actuated signal settings. Fixed-time signal settings can be viewed as an instance of actuated signal settings. Hence, the proposed approach holds for fixed-time signal settings. Our goal is to define a formulation that depends on average (over time) signal setting parameters, such as cycle length, green or red phase duration, or green split. In this paper, we propose a formulation that only relies on knowledge of the (average) green split, denoted $g$, which is the average ratio of the green phase duration to the cycle length. Hereafter all signal plan parameters are time-averaged.

The proposed formulation assumes both $\beta$ and the ratio $\frac{\beta}{\alpha}$ to be linear functions of the green split $g$, i.e.,
\begin{subnumcases}{}
v = v_{max} \left( 1- \left(\frac{q}{q_{cap}} \right)^\alpha \right)^\beta \label{eq:s_fd}\\
\beta = \theta_0 + \theta_1 g    \label{eq:beta_reg}\\ 
\frac{\beta}{\alpha} = \theta_2 + \theta_3 g, \label{eq:beta_alpha_reg}
\end{subnumcases}

\noindent where the new model parameters are $\theta_0, \theta_1, \theta_2$, and $\theta_3$. 

These parameters are fitted for all signalized segments of a given city. This means that all parameters of the proposed FD formulation can be readily estimated without the need to collect extensive segment-specific data. In other words, compared to the previous formulation, the parameters $\alpha$ and $\beta$ no longer require segment-specific data to be estimated. This makes the proposed formulation more scalable and more broadly applicable. 

As our numerical validation experiments with Salt Lake City (Utah, USA) data indicate, although the expression considers the time-average green spilt, it holds for actuated signals where the green split can vary substantially over time. For fixed-time or pre-timed signal settings, the average would represent an average over time-of-day-specific constant green splits, since the green splits typically vary over the course of the day.

\section{Results and discussion}
We validate the proposed formulation with the Salt Lake City (Utah, USA) data of actuated signal controlled urban road segments. We present the data first, and then we present representative results. 

\subsection{Data}
We consider actuated signalized intersections. We focus on road segments that are adjacent to the signalized intersections and correspond to the major traffic movement directions (i.e., protected signal phases 2 \& 6). %However, the operations during daytime (7AM - 8PM) during weekdays approximate the pre-time signal control given the signals are semi-actuated with bounds on the cycle length. 
The definition of major streets and detailed signal event data is obtained from the Utah Department of Transportation (UDOT) Automated Traffic Signal Performance Metrics (ATSPM) \citep{udot_atspm}. Through the ATSPM portal, we have access to lane-level vehicular count (i.e., volume or flow) data obtained at stop-bar detectors. We aggregate counts at the hourly and the segment level. The flow values are segmented into intervals of 30 veh/hr-lane. 

We use aggregated and anonymized space-mean speeds of either a given segment (all traffic movements combined) or per traffic movement. We carry out an average analysis over space and/or time, meaning that for actuated signals we focus on the average (over time) signal settings (namely cycle, green, and red time). All data, parameters, and variables mentioned hereafter are assumed to be averages over time or over space.

% The segment configurations such as the maximum speeds and the number of lanes are obtained from the Google Maps.

We consider segments that have similar average cycle length values for 7AM - 8PM of weekdays (Mon. - Fri.).
The cycle length values considered are  $+$/$-$ 5s of 114s. 

% both (1) the ground truth data based on which the FD is derived, and (2) the FD curve derived based on Equation~\eqref{eq:FD1}. 

\subsection{Analysis}
\begin{figure}[ht]
	\center
	\includegraphics[width=0.85\textwidth]{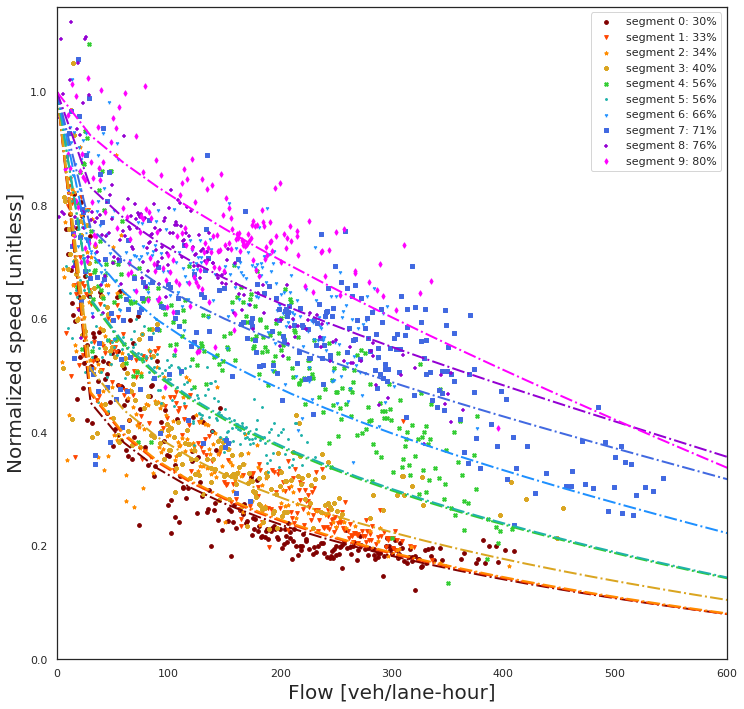}
	\caption{Comparison of signal-parametrized FDs and real data.}
    \label{fig:s_fd_analysis}
\end{figure}

%We present preliminary results on the possibility of parameterizing FD by signal plans. 
Figure~\ref{fig:s_fd_analysis} considers a set of 10 segments that represent a wide range of signal settings. It displays the FDs of these segments. Each segment is displayed in a different color. The $x$-axis represents segment flows (in the unit of veh/lane-hour) and the $y$-axis displays the corresponding speed normalized by the speed limit. These 10 segments are selected to represent a wide range of the associated green split values, ranging from 0.3 (i.e., the green phase is active 30\% of the cycle time) to 0.8 (i.e., the green phase is active 80\% of the cycle time). For each segment, a dash-dotted curve, of the same color as the segment, represents the corresponding signal-parametrized FD curve defined by Equations~\eqref{eq:s_fd}-\eqref{eq:beta_alpha_reg}.

%For all cases, the signal-parametrized FD curves show a one-to-one mapping for flow-speed in an urban signalized controlled network. 
This plot indicates that the proposed formulation describes well how the FDs vary as a function of the green split. As the green split increases, the FDs are shifted higher up. In other words, for a given $x$ value of flow the corresponding speed (i.e. $y$ value) increases with the green split. This is consistent with the intuition that an increased green split, corresponding to an increased flow capacity, allows speeds to decay from their free-flowing values at higher flow values. Compared to the work of \cite{li22}, the proposed formulation adds interpretability to the power coefficients of the FD. This is of practical relevance and can serve to estimate the impact of signal setting changes on congestion patterns. 
% to enhance traffic management strategies and
% \newpage

\section{Conclusion and discussion}
We postulate a parsimonious parametric urban FD for signalized segments with power coefficients that are explicit functions of the green split of the underlying signal plan. The formulation is validated using the Salt Lake City data. The experiments show that it is possible to parameterize FD functional forms by the signal plans and the derived traffic signal-parametrized FD closely approximates the ground truth. 

The parameters of the proposed linear formulations (the $\theta$ scalars) that define the power coefficients of the FD have common values across segments. This makes the model less reliant on segment-specific data. Hence, it is more scalable and can be more readily applied across segments. It is an open question whether the numerical values found for these parameters hold across cities. 

% (3) this research serves as the stepping stone for many applications related to traffic state estimation, traffic management, and transportation planning.

This is a part of the ongoing effort. The most recent learnings and formulations will be presented at the conference. Some directions for further research include: (1) understanding the impact of signal coordination (e.g., offsets) on the FDs; and (2) incorporating other factors (e.g., left turning flows) into the FD formulation to further enhance our understanding of what governs the shape of the FDs. 

\bibliography{biblio_spacetime}
 
\end{document}